\documentclass[aps,prc,superscriptaddress,twoside,twocolumn,nofootinbib,10pt,%
showpacs,floatfix]{revtex4-1}

\usepackage{amsmath,amssymb}
\usepackage{graphicx,bm}
\usepackage{epstopdf}
\usepackage{ulem} 
\usepackage[usenames]{color}
\usepackage{float}

\allowdisplaybreaks

\newcommand{\non}{\nonumber}
\newcommand{\del}{\partial}
\newcommand{\mpi}{m_{\pi}}

\begin{document}

\title{
Heavy quark spin multiplet structure of $P_c$-like pentaquark as P-wave hadronic molecular state
}

\author{Yuki Shimizu}
\email{yshimizu@hken.phys.nagoya-u.ac.jp}
\affiliation{Department of Physics,  Nagoya University, Nagoya 464-8602, Japan}

\author{Yasuhiro Yamaguchi}
\email{yasuhiro.yamaguchi@riken.jp}
\affiliation{Theoretical Research Division, Nishina Center, RIKEN, Hirosawa, Wako, Saitama 351-0198, Japan}

\author{Masayasu Harada}
\email{harada@hken.phys.nagoya-u.ac.jp}
\affiliation{Department of Physics,  Nagoya University, Nagoya 464-8602, Japan}

\date{\today}


\begin{abstract}
We study the heavy quark spin (HQS) multiplet structure of P-wave $Q\bar{Q}qqq$-type 
pentaquarks treated as molecules of a heavy meson and a heavy baryon.
We define the light-cloud spin (LCS) basis
decomposing the meson-baryon spin wavefunction into the LCS and HQS parts.
 Introducing the LCS basis, we find HQS multiplets classified by
 the LCS; five HQS singlets, two HQS doublets, and three HQS triplets.
We construct the one-pion exchange potential respecting 
the heavy quark spin and chiral symmetries
 to demonstrate which HQS multiplets are realized as a bound state.
By solving the coupled channel Schr\"odinger equations, we study
the heavy meson-baryon systems
with $I=1/2$ and $J^P=(1/2^+, 3/2^+, 5/2^+, 7/2^+)$.
The bound states which have same LCS 
structure are degenerate at the heavy quark limit,
and the degeneracy is resolved for finite mass. 
This HQS multiplet structure will be measured in the future experiments.
\end{abstract}

\maketitle


\section{Introduction}
\label{sec:Intro}
In 2015, the Large Hadron Collider beauty experiment (LHCb) collaboration 
observed two hidden charm pentaquarks, $P_c^+(4380)$ and $P_c^+(4450)$
\cite{Aaij:2015tga, Aaij:2016phn, Aaij:2016ymb}.
Their masses are $M_{P_c(4380)} = 4380\pm8\pm28$ MeV and $M_{P_c(4450)} = 4449.8\pm1.7\pm2.5$ MeV, 
and decay widths are $\Gamma_{P_c(4380)} = 205\pm18\pm86$ MeV and $\Gamma_{P_c(4450)} = 39\pm5\pm19$ MeV.
Their spin and parity are not determined.
The one state has $J=3/2$ and the other has $J=5/2$ and their parity is opposite.

The $P_c$ pentaquarks have a charm quark and an anti-charm quark.
They are called the hidden-charm pentaquarks.
There were some theoretical works of hidden-charm pentaquarks
before the LHCb announcement \cite{Wu:2010jy, Yang:2011wz, Wang:2011rga, Wu:2012md}.
After the LHCb observation, many theoretical studies in various ways have been conducted: 
hadronic molecular picture 
\cite{Chen:2015loa, He:2015cea, Chen:2015moa, Huang:2015uda, Roca:2015dva, Meissner:2015mza, Xiao:2015fia, Burns:2015dwa, Kahana:2015tkb, Chen:2016otp, Shimizu:2016rrd, Yamaguchi:2016ote, He:2016pfa, Ortega:2016syt, Azizi:2016dhy, Geng:2017hxc},
quark model estimation \cite{Santopinto:2016pkp, Wu:2017weo, Hiyama:2018ukv},
diquark picture \cite{Maiani:2015vwa, Lebed:2015tna, Li:2015gta, Wang:2015epa, Zhu:2015bba},
quark-cluster model\cite{Takeuchi:2016ejt},
baryocharmonium model\cite{Kubarovsky:2015aaa}, 
hadroquarkonia model\cite{Eides:2017xnt}, 
soliton model\cite{Scoccola:2015nia}, 
holographic QCD \cite{Liu:2017frj},
and hadronic molecule coupled with five-quark state \cite{Yamaguchi:2017zmn}.
Some review papers are also published \cite{Chen:2016qju, Ali:2017jda, Guo:2017jvc}.

The hadronic molecular picture is one of the highly possible model around the hadron threshold.
The threshold of $\bar{D}\Sigma_c^*$ is $4385.3$ MeV and 
$\bar{D}^*\Sigma_c$ is $4462.2$ MeV.
These values are slightly above the mass of $P_c(4380)$ and $P_c(4450)$, respectively.
Therefore, the $P_c$ pentaquarks can be considered as the loosely bound states of a charmed meson and a charmed baryon.

In the heavy quark effective theory, the spin dependent interaction of
heavy quark is suppressed by the inverse of the heavy quark mass,
$1/m_{Q}$.
At the heavy quark limit, therefore, the dynamics is independent of
the transformation of the heavy quark spin. This is called the heavy
quark spin symmetry (HQSS).
The suppression of the spin dependent force causes the
decomposition of the heavy quark spin
and the light-cloud spin at the heavy quark limit \cite{Isgur:1989vq, Isgur:1989ed, Isgur:1991wq, Neubert:1993mb, Manohar:2000dt} : 
\begin{align}
\vec{J} = \vec{s}_{\rm light} + \vec{s}_{\rm heavy}~.
\label{eq:Heavy Quark Symmetry}
\end{align}
The total angular momentum $\vec{J}$ is a conserved quantity, 
and the heavy quark spin $\vec{s}_{\rm heavy}$ is conserved at heavy quark limit.
Then, the light-cloud spin $\vec{s}_{\rm light}$ is also conserved.

The HQSS leads to the mass degeneracy between the heavy hadrons with different spin.
Considering a heavy meson $P^{(*)} \sim Q\bar{q}$ 
with a heavy quark $Q$ and an anti-light quark $\bar{q}$,
the total spin of $Q\bar{q}$ is
\begin{align}
{J}_{\pm} = 1/2 \pm 1/2~,
\end{align}
and $J_+=1$ ($J_-=0$) for the vector meson $P^\ast$ (the pseudoscalar meson $P$),
because of the quark spin $1/2$.
Their difference comes from the spin configuration of the light cloud and heavy quark spins.
However, the system is independent of the heavy quark spin and as a result, 
the spin $0$ state $P$ and the spin $1$ state $P^\ast$ degenerate at heavy quark limit.
This structure is called the HQS doublet.

In the real world, however, the quark masses are finite, 
so that there exists a mass difference between the pseudoscalar and vector mesons.
For example, the mass difference between pseudoscalar meson $K$ and vector meson $K^*$ is about $400$ MeV.
By contrast, 
the mass splitting between $D$ and $D^*$ is about $140$MeV and 
between $B$ and $B^*$ is $45$MeV.
The mass difference is much smaller in the charm and bottom quark sectors than in the light quark sector.
There is a same tendnecy in the single heavy baryon.
The mass difference between spin $1/2$ baryon $\Sigma_c$ and spin $3/2$ baryon $\Sigma_c^*$ is about $65$ MeV, and $\Sigma_b$ and $\Sigma_b^*$ is about $20$ MeV.
The HQSS approximately exists in the heavy quark sector.

The purpose of this work is to study the HQS multiplet structure of 
$Q\bar{Q}qqq$-type pentaquarks as 
hadronic molecular states of a $\bar{P}^{(*)}$ meson and a $\Sigma_Q^{(*)}$ baryon.
Here $\bar{P}$ and $\bar{P}^*$ denote mesons with $J^P = 0^-$ and $1^-$ 
with an anti-heavy quark like $\bar{D}$ and $\bar{D}^*$ mesons, and 
$\Sigma_{Q}$ and $\Sigma_Q^*$ the baryons with $J^P =1/2^+$ and $3/2^+$
with a heavy quark like $\Sigma_c$ and $\Sigma_c^*$ baryons.
We note that the HQS doublet  structure of single heavy hadrons such as 
$\bar{P}^{(*)}$ and $\Sigma_Q^{(*)}$ is well known.
Furthermore, the HQS multiplet structure of 
multi-hadron system with single heavy quark like $\bar{P}^{(*)}N$
molecular states has been studied
in Refs.~\cite{Yasui:2013vca, Yamaguchi:2014era, Hosaka:2016ypm}.
On the other hand, the HQS multiplet structure of doubly heavy hadrons 
like $Q\bar{Q}qqq$ pentaquarks 
is nontrivial.
Hence, it is interesting to 
investigate the HQS multiplet structure of heavy meson-baryon molecular states.

In Ref.~\cite{Shimizu:2018ran},
the HQS multiplet structure of $\bar{P}^{(*)}\Sigma_Q^{(*)}$ with S-wave has been studied.
The analysis of the S-wave state covers the
negative parity Pentaquarks.
However, one of the $P_c^+$ pentaquarks has the positive parity.
Accordingly, we study possible positive parity states by considering 
P-wave molecular states of $\bar{P}^{(*)}$ and $\Sigma_Q^{(*)}$ 
in the present paper.
We define the light-cloud spin (LCS) basis introduced in Ref.~\cite{Shimizu:2018ran}
which is useful to investigate the HQS multiplet structure.
Although the hadronic molecular (HM) basis
is simple to consider the possible total spin state 
and to construct the potentials, it is not useful to investigate the HQS multiplet structure
because the heavy quark spin and the light-cloud spin are not separated
in this basis.
In the light-cloud spin (LCS) basis, however,
these spins are separated explicitly as shown in Eq.~\eqref{eq:Heavy Quark Symmetry}.
Below we shall transform the HM basis to the LCS basis and study the HQS multiplet structure. 
Moreover, we demonstrate which multiplets can be bound under the one-pion exchange potential (OPEP) from the heavy hadron effective theory.

This paper is organized as follows.
In Sec.~\ref{sec:HQS multiplet Pwave} we construct the HM basis of the $\bar{P}^{(\ast)}\Sigma^{(\ast)}_{Q}$ states
and transfer it to LCS basis to discuss the HQS multiplet structure.
The OPEP is shown in Sec.~\ref{sec:Lagrangian and Potential}.
In Sec.~\ref{sec:Numerical result}, 
we summarize the numerical results.
Finally, Sec.~\ref{sec:summary} is a summary and discussion.

\section{HQS multiplet structure of $\bar{P}^{(*)}\Sigma_Q^{(*)}$ with P-wave}
\label{sec:HQS multiplet Pwave}
In this section, we consider the HQS multiplet of P-wave states.
First, we construct the hadronic molecular
(HM) basis of the $\bar{P}^{(*)}\Sigma_Q^{(*)}$ states.
The possible spin states and meson-baryon components with given $J^P$ are shown in Tab.~\ref{tab:possible spin state}.
\begin{table*}[!htbp]
\begin{center}
\caption{Possible spin states of
 the P-wave $\bar{P}^{(*)}\Sigma_Q^{(*)}$ molecular states with given $J^P$.
$\left(^{2S+1}L_{J}\right)$ denotes the total spin of meson and baryon $S$, the orbital angular momentum $L$, and the total angular momentum $J$.
}
\begin{tabular}{c|l}\hline
 $J^P$ &  \\ \hline
 $\frac{1}{2}^+$ & $\bar{P}\Sigma_{Q}\left( ^2P_{1/2} \right), \bar{P}\Sigma_{Q}^*\left( ^4P_{1/2} \right), \bar{P}^*\Sigma_{Q}\left( ^2P_{1/2} \right), \bar{P}^*\Sigma_{Q}\left( ^4P_{1/2} \right), \bar{P}^*\Sigma_{Q}^*\left( ^2P_{1/2} \right), \bar{P}^*\Sigma_{Q}^*\left( ^4P_{1/2} \right)$  \\[1mm] \hline
 $\frac{3}{2}^+$ & $\bar{P}\Sigma_{Q}\left( ^2P_{3/2} \right), \bar{P}\Sigma_{Q}^*\left( ^4P_{3/2} \right), \bar{P}^*\Sigma_{Q}\left( ^2P_{3/2} \right), \bar{P}^*\Sigma_{Q}\left( ^4P_{3/2} \right), \bar{P}^*\Sigma_{Q}^*\left( ^2P_{3/2} \right), \bar{P}^*\Sigma_{Q}^*\left( ^4P_{3/2} \right), \bar{P}^*\Sigma_{Q}^*\left( ^6P_{3/2} \right)$  \\[1mm] \hline
 $\frac{5}{2}^+$ & $\bar{P}\Sigma_{Q}^*\left( ^4P_{5/2} \right), \bar{P}^*\Sigma_{Q}\left( ^4P_{5/2} \right), \bar{P}^*\Sigma_{Q}^*\left( ^4P_{5/2} \right), \bar{P}^*\Sigma_{Q}^*\left( ^6P_{5/2} \right)$  \\[1mm] \hline
 $\frac{7}{2}^+$ & $\bar{P}^*\Sigma_{Q}^*\left( ^6P_{7/2} \right)$  \\[1mm] \hline
\end{tabular}
\label{tab:possible spin state}
\end{center}
\end{table*}
Giving the $\bar{P}^{(\ast)}\Sigma^{(\ast)}_{Q}$ component with total spin $S$, 
we obtain the spin structure in the HM basis and the possible total angular momentum $J$
 as follows:
\begin{align}
	\bar{P}\Sigma_Q\left(^2P\right) &= 
 \left[ P \left[ \left[\bar{Q}q\right]_0
 \left[Q\left[ d \right]_1\right]_{1/2} \right]_{1/2} \right] = \frac{1}{2} \oplus \frac{3}{2}, \\
	\bar{P}\Sigma_Q^*\left(^4P\right) &= 
 \left[ P \left[ \left[\bar{Q}q\right]_0
 \left[Q\left[ d \right]_1\right]_{3/2} \right]_{3/2} \right] = \frac{1}{2} \oplus \frac{3}{2} \oplus \frac{5}{2}, \\
	\bar{P}^*\Sigma_Q\left(^2P\right) &= 
 \left[ P \left[ \left[\bar{Q}q\right]_1 
 \left[Q\left[ d \right]_1\right]_{1/2} \right]_{1/2} \right] = \frac{1}{2} \oplus \frac{3}{2}, \\
	\bar{P}^*\Sigma_Q\left(^4P\right) &= 
 \left[ P \left[ \left[\bar{Q}q\right]_1 
 \left[Q\left[ d \right]_1\right]_{1/2} \right]_{3/2} \right] = \frac{1}{2} \oplus \frac{3}{2} \oplus \frac{5}{2}, \\
	\bar{P}^*\Sigma_Q^*\left(^2P\right) &= 
 \left[ P \left[ \left[\bar{Q}q\right]_1 
 \left[Q\left[ d \right]_1\right]_{3/2} \right]_{1/2} \right] = \frac{1}{2} \oplus \frac{3}{2}, \\
	\bar{P}^*\Sigma_Q^*\left(^4P\right) &= 
 \left[ P \left[ \left[\bar{Q}q\right]_1 
 \left[Q\left[ d \right]_1\right]_{3/2} \right]_{3/2} \right] = \frac{1}{2} \oplus \frac{3}{2} \oplus \frac{5}{2}, \\
	\bar{P}^*\Sigma_Q^*\left(^6P\right) &= 
 \left[ P \left[ \left[\bar{Q}q\right]_1 
 \left[Q\left[ d \right]_1\right]_{3/2} \right]_{5/2} \right] = \frac{3}{2} \oplus \frac{5}{2} \oplus \frac{7}{2} \ ,
\end{align}
where $\left[ L \left[ \left[ \bar{Q}q \right]_{s_1}\left[ Q[ d ]_{1} \right]_{s_2}  \right]_{S} \right]$ 
implies that the $\bar{P}^{(*)} \sim \bar{Q}q$ meson with spin $s_1$ and  
the $\Sigma_{Q}^{(*)} \sim Qd$ baryon ($d=qq$ is a light diquark) with spin $s_2$ are combined into
a $\bar{P}^{(*)}\Sigma_{Q}^{(*)}$ composite state with 
the total spin $S$ and the orbital angular momentum $L$.
$[d]_{1}$ implies that the spin of the diquark is $1$.
HM basis is simple to construct the possible spin states because it is just the 
coupling of the spins of the meson and baryon 
and the orbital angular momentum.
However, HM basis is not suitable to discuss the HQS multiplet structure.
The heavy quark spin and the light-cloud spin are independently conserved in the heavy quark limit.
Thereby, the heavy quark spin and the other spin must be treated separately.
We define the light-cloud spin (LCS) basis as a suitable basis to study the structure of HQS multiplets.

In the LCS basis, the spin structures are rewritten as follows : 
\begin{align}
& \mbox{(s-1) : }
\left[ \left[\bar{Q}Q\right]_0 \left[ P\left[q\left[d\right]_1\right]_{1/2} \right]_{1/2} \right] = \frac{1}{2} \ ,\label{LCS s-1} \\ 
& \mbox{(s-2) : } 
\left[ \left[\bar{Q}Q\right]_0 \left[ P\left[q\left[d\right]_1\right]_{3/2} \right]_{1/2} \right] = \frac{1}{2} \ , \label{LCS s-2} \\
& \mbox{(s-3) : }
\left[ \left[\bar{Q}Q\right]_0 \left[ P\left[q\left[d\right]_1\right]_{1/2} \right]_{3/2} \right] = \frac{3}{2} \ , \label{LCS s-3}\\ 
& \mbox{(s-4) : }
\left[ \left[\bar{Q}Q\right]_0 \left[ P\left[q\left[d\right]_1\right]_{3/2} \right]_{3/2} \right] = \frac{3}{2}\ , \label{LCS s-4} \\ 
& \mbox{(s-5) : }
\left[ \left[\bar{Q}Q\right]_0 \left[ P\left[q\left[d\right]_1\right]_{3/2} \right]_{5/2} \right] = \frac{5}{2} \ , \label{LCS s-5} \\ 
& \mbox{(d-1) : }
\left[ \left[\bar{Q}Q\right]_1 \left[ P\left[q\left[d\right]_1\right]_{1/2} \right]_{1/2} \right] = \frac{1}{2} \oplus \frac{3}{2} \ , \label{LCS d-1} \\ 
& \mbox{(d-2) : }
\left[ \left[\bar{Q}Q\right]_1 \left[ P\left[q\left[d\right]_1\right]_{3/2} \right]_{1/2} \right] = \frac{1}{2} \oplus \frac{3}{2} \ , \label{LCS d-2} \\ 
& \mbox{(t-1) : }
\left[ \left[\bar{Q}Q\right]_1 \left[ P\left[q\left[d\right]_1\right]_{1/2} \right]_{3/2} \right] = \frac{1}{2} \oplus \frac{3}{2} \oplus \frac{5}{2} \ , \label{LCS t-1} \\ 
& \mbox{(t-2) : }
\left[ \left[\bar{Q}Q\right]_1 \left[ P\left[q\left[d\right]_1\right]_{3/2} \right]_{3/2} \right] = \frac{1}{2} \oplus \frac{3}{2} \oplus \frac{5}{2} \ , \label{LCS t-2} \\ 
& \mbox{(t-3) : }
\left[ \left[\bar{Q}Q\right]_1 \left[ P\left[q\left[d\right]_1\right]_{3/2} \right]_{5/2} \right] = \frac{3}{2} \oplus \frac{5}{2} \oplus \frac{7}{2} \ ,\label{LCS t-3} 
\end{align}
where $\left[\bar{Q}Q\right]_{s_1} \left[ P\left[q\left[d\right]_1\right]_{s_2} \right]_{s_3}$ implies the followings: A heavy quark $Q$ and an anti-heavy quark $\bar{Q}$ are combined into a state with spin $s_1$ in S-wave. Spins of a light quark $q$ and a diquark $d$ are coupled to spin $s_2$ and the total spin of the combined state in P-wave is given by $s_3$.
The right hand side of the equation shows the possible spins of the combined pentaquark states.
There exist five HQS singlets (s-1 to s-5), two HQS doublets (d-1 and d-2), and three HQS triplets (t-1 to t-3).
The HQS triplet does not exist in single heavy hadrons.
It is a feature of the multi-heavy quark system.
The basis transformation is done by 
\begin{align}
\psi_{J^P}^{\rm LCS} = U_{J^P}^{-1}\psi_{J^P}^{\rm HM},
\label{eq:wave function transformation}
\end{align}
where the $U$ is a transformation matrix determined by the Clebsch-Gordan coefficient to reconstruct the spin structure.
The detail of the basis transformation is summarized in Appendix~\ref{sec:Basis transformation}.
It is to be noted that the two heavy quarks are labeled by the same velocity $v$ to 
classify the pentaquark states based on the heavy quark spin symmetry.


\section{Potentials}
\label{sec:Lagrangian and Potential}
In the previous section, we showed 
that there are ten multiplets in the P-wave $\bar{P}^{(*)}\Sigma_Q^{(*)}$ molecular states.
In this section, we demonstrate that which of the multiplets can be bound by using one-pion exchange potential (OPEP).
We construct the OPEP for $\bar{P}^{(*)}\Sigma_Q^{(*)}$ molecular states based on the heavy hadron effective theory.

The $\bar{P}^{(*)}$ meson and pion interaction Lagrangian is given in Refs.\cite{Falk:1991nq, Wise:1992hn, Cho:1992gg, Yan:1992gz, Falk:1992cx},
and the $\Sigma_Q^{(*)}$ baryon and pion interaction Lagrangian is given in Refs.\cite{Yan:1992gz, Liu:2011xc}.
See also our previous paper \cite{Shimizu:2018ran}.

When we construct the OPEP from effective Lagrangians,
we introduce a cutoff parameter $\Lambda$
via the monopole type form factor
\begin{align}
	F(q) = \frac{\Lambda^2 - m_{\pi}^2}{\Lambda^2 + |\vec{q}\,|^2}~,
\end{align}
at each vertex, where $m_{\pi}$ is the mass of the exchanging pion, and $\vec{q}$ is its momentum.
We use the same cutoff for $\bar{P}^{(*)}\bar{P}^{(*)}\pi$ and $\Sigma_Q^{(*)}\Sigma_Q^{(*)}\pi$ vertices for simplicity, 
and fix the value of cutoff $1000$ MeV and $1500$ MeV. 
The obtained potential matrices in the HM basis are summarized in Appendix \ref{sec:Basis transformation}.
We note that contact terms are subtracted from the potentials, 
because in a conventional way, the OPEP has been considered at large distance~\cite{Bohr-Mottelson}.
Furthermore, we study the cases where the final pentaquarks carry isospin $1/2$, 
because $P_c$ pentaquarks carry $I=1/2$.

\begin{figure}[!tbp]
\begin{center}
\includegraphics[bb = 0 0 640 384, width=0.45\textwidth]{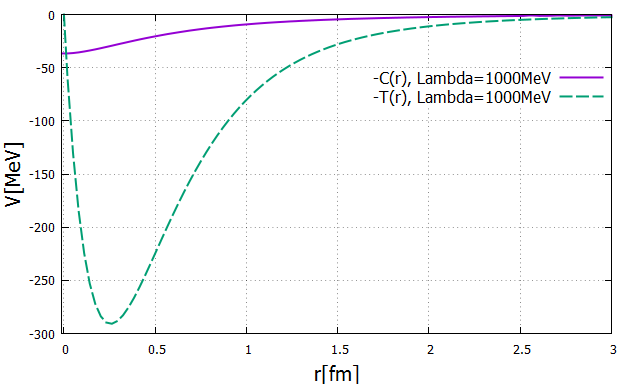}
\caption{
Spin-spin potential $C$ and tensor potential $T$ with the cutoff parameter $\Lambda=1000$ MeV.
}
\label{fig:OPEP_L1000}
\end{center}
\end{figure}

The potential matrices can be also transformed to LCS basis by using the unitary matrix $U$ as follows: 
\begin{widetext}
\begin{align}
V_{1/2^+}^{\rm LCS} &= U_{1/2^+}^{-1} V_{1/2^+}^{\rm HM} U_{1/2^+} \non \\
&= \left(
\begin{array}{cccccc}
	C & -\frac{\sqrt{2}}{2}T & 0 & 0 & 0 & 0 \\
	-\frac{\sqrt{2}}{2}T & -\frac{1}{2}C + T & 0 & 0 & 0 & 0 \\
	0 & 0 & C & -\frac{\sqrt{2}}{2}T & 0 & 0 \\
	0 & 0 & -\frac{\sqrt{2}}{2}T & -\frac{1}{2}C + T & 0 & 0 \\
	0 & 0 & 0 & 0 & C & \frac{\sqrt{5}}{10}T \\
	0 & 0 & 0 & 0 & \frac{\sqrt{5}}{10}T & -\frac{1}{2}C - \frac{4}{5}T
\end{array}
\right)\frac{gg_1}{f_{\pi}^2}~,
\label{eq:potential in LCS 1/2}
\end{align}
\begin{align}
V_{3/2^+}^{\rm LCS} &= U_{3/2^+}^{-1}V_{3/2^+}^{\rm HM}U_{3/2^+} \non \\
&= \left(
	\begin{array}{ccccccc}
		C & \frac{\sqrt{5}}{10}T & 0 & 0 & 0 & 0 & 0 \\
		\frac{\sqrt{5}}{10}T & -\frac{1}{2}C-\frac{4}{5}T & 0 & 0 & 0 & 0 & 0 \\
		0 & 0 & C & -\frac{\sqrt{2}}{2}T & 0 & 0 & 0 \\
		0 & 0 & -\frac{\sqrt{2}}{2}T & -\frac{1}{2}C + T & 0 & 0 & 0 \\
		0 & 0 & 0 & 0 & C & \frac{\sqrt{5}}{10}T & 0 \\
		0 & 0 & 0 & 0 & \frac{\sqrt{5}}{10}T & -\frac{1}{2}C - \frac{4}{5}T & 0 \\
		0 & 0 & 0 & 0 & 0 & 0 & -\frac{1}{2}C + \frac{1}{5}T
	\end{array}
\right)\frac{gg_1}{f_{\pi}^2}~,
\label{eq:potential in LCS 3/2}
\end{align}
\begin{align}
V^{\rm LCS}_{5/2^+} &= U_{5/2^+}^{-1} V^{\rm HM}_{5/2^+} U_{5/2^+} \non \\
&\hspace{-5pt}= \left(
	\begin{array}{cccc}
		-\frac{1}{2}C + \frac{1}{5}T & 0 & 0 & 0 \\
		0 & C & \frac{\sqrt{5}}{10}T & 0 \\
		0 & \frac{\sqrt{5}}{10}T & -\frac{1}{2}C - \frac{4}{5}T & 0 \\
		0 & 0 & 0 & -\frac{1}{2}C + \frac{1}{5}T
	\end{array}
\right)\frac{gg_1}{f_{\pi}^2}~,
\label{eq:potential in LCS 5/2}
\end{align}
\end{widetext}
\begin{align}
V^{\rm LCS}_{7/2^+} &= U_{7/2^+}^{-1} V^{\rm HM}_{7/2^+} U_{7/2^+} \non \\
&= \frac{gg_1}{f_{\pi}^2}\left[ -\frac{1}{2}C + \frac{1}{5}T \right]~,
\label{eq:potential in LCS 7/2}
\end{align}
where the functions $C(\mpi, r, \Lambda)$ and $T(\mpi, r, \Lambda)$ are the spin-spin potential and tensor potential, respectively.
We omitted the arguments of potentials in the above equations.
Their explicit forms are given by
\begin{align}
	C(\mpi, r, \Lambda) &= \frac{m_{\pi}^2}{4\pi}\left[ \frac{e^{-m_{\pi}r} - e^{-\Lambda r}}{r} - \frac{\Lambda^2 - m_{\pi}^2}{2\Lambda}e^{-\Lambda r} \right]~, \\
	T(\mpi, r, \Lambda) &= \frac{m_{\pi}^3}{4\pi}H_3(m_{\pi}, \Lambda, r)~, \\
	H_3(m_{\pi}, \Lambda, r) &= \frac{1}{m_{\pi}^3}
	\left[ \frac{m_{\pi}^2r^2 + 3m_{\pi}r + 3}{r^3}e^{-m_{\pi}r} \right. \non\\
	&\hspace{25pt}\left. -\frac{\Lambda^2r^2 + 3\Lambda r + 3}{r^3}e^{-\Lambda r} \right. \non\\
	&\hspace{25pt}\left. -\frac{\Lambda^2 - m_{\pi}^2}{2r}e^{-\Lambda r} 
	- \frac{\Lambda^3 - \Lambda m_{\pi}^2}{2}e^{-\Lambda r} \right]~.
\end{align}
The typical shapes of $C$ and $T$ are shown in Fig.\ref{fig:OPEP_L1000}.
This shows that the signs $C$ and $T$ are negative so as to be attractive potentials.
The heavy meson - pion coupling constant $|g| = 0.59$ is determined by the decay of $D^* \to D\pi$ \cite{Olive:2016xmw}, 
the heavy baryon - pion coupling constant $g_1 = 0.94$ is estimated by the quark model in Ref.\cite{Liu:2011xc},
and the pion decay constant is $f_{\pi} = 92.4$ MeV.

The block matrices of the above OPEPs in LCS basis are classified by the HQS multiplet structure.
For example, the first $2\times2$ block in Eq.(\ref{eq:potential in LCS 1/2})
is for HQS singlet sector with the total spin $1/2$.
It corresponds to the first and second component in Eq.(\ref{eq:wave function in LCS 1/2}), or (s-1) component in Eq.(\ref{LCS s-1}) and (s-2) in Eq.(\ref{LCS s-2}).
Similarly, the second block in Eq.(\ref{eq:potential in LCS 1/2}) corresponds to 
the third and forth component in Eq.(\ref{eq:wave function in LCS 1/2})
which is for spin $(1/2, 3/2)$ doublet,
and the third block in Eq.(\ref{eq:potential in LCS 1/2}) corresponds to 
the fifth and sixth component in Eq.(\ref{eq:wave function in LCS 1/2})
which is for spin $(1/2, 3/2, 5/2)$ triplet.

The component of block matrix is determined by the structure of the light-cloud spin.
For instance, the first and second block in Eq.(\ref{eq:potential in LCS 1/2}) are identical 
because 
two $1/2$ singlets and two $(1/2, 3/2)$ doublets have the same light-cloud spin structure, 
$\left[ P\left[q\left[d\right]_1\right]_{1/2} \right]_{1/2}$ and 
$\left[ P\left[q\left[d\right]_1\right]_{3/2} \right]_{1/2}$,
as shown in Eq.(\ref{eq:wave function in LCS 1/2}).


\section{Numerical result}
\label{sec:Numerical result}
In this section we show the binding energy obtained by solving the coupled channel Schr\"odinger equation under the  OPEPs obtained in the previous section.
We use the Gaussian expansion method \cite{Hiyama:2003cu} 
to solve the Schr\"odinger equations.
As discussed in Ref.\cite{Shimizu:2018ran}, the coupling constant has a sign ambiguity.
This ambiguity is the relative sign between the heavy meson - pion coupling $g$ and
the heavy baryon - pion coupling $g_1$.
We assign the same sign as the quark model estimation in Ref.\cite{Liu:2011xc}
as the sign of $g_1$.
Therefore, we treat the sign uncertainty as the sign of $g$.
In this study, we investigate both cases.

We include the effect of the heavy quark spin symmetry breaking 
by introducing the mass difference between two heavy mesons (baryons) in one HQS doublet,
namely $\bar{P}$ and $\bar{P}^*$ ($\Sigma_Q$ and $\Sigma_Q^*$).
We parameterize the heavy hadron masses as done 
in Ref.\cite{Shimizu:2018ran} : 
\begin{align}
M_{\bar{P}} &= 2\mu + \frac{a}{2\mu} + \frac{w}{(2\mu)^2}~, 
\label{eq:mass fit 0meson} \\
M_{\bar{P}^*} &= 2\mu + \frac{b}{2\mu} + \frac{x}{(2\mu)^2}~, 
\label{eq:mass fit 1meson} \\
M_{\Sigma_Q} &= 2\mu + \frac{c}{2\mu} + \frac{y}{(2\mu)^2}~, 
\label{eq:mass fit 12baryon} \\
M_{\Sigma_Q^*} &= 2\mu + \frac{d}{2\mu} + \frac{z}{(2\mu)^2}~. 
\label{eq:mass fit 32baryon} 
\end{align}
The mass parameter $\mu$ controls the typical mass scale.
It corresponds to the averaged reduced mass of 
$\bar{P}\Sigma_Q, \bar{P}\Sigma_Q^*, \bar{P}^*\Sigma_Q$, and $\bar{P}^*\Sigma_Q^*$.
We determine the eight parameters $a,b,c,d,w,x,y$, and $z$ to reproduce the eight hadron masses shown in Table \ref{tab:hadron masses}.
\begin{table}[btp]
\centering
\caption{Masses 
of relevant charmed and bottomed 
hadrons~\cite{Olive:2016xmw}.}
\begin{tabular}{c|cccc}\hline
	 & $\bar{D}$ & $\bar{D}^*$ & $B$ & $B^*$ \\
	Mass[MeV] & $1867.21$ & $2008.56$ & $5279.48$ & $5324.65$ \\ \hline 
	 & $\Sigma_c$ & $\Sigma_c^*$ & $\Sigma_b$ & $\Sigma_b^*$ \\
	Mass[MeV] & $2453.54$ & $2518.13$ & $5813.4$ & $5833.6$ \\ \hline 
\end{tabular}
\label{tab:hadron masses}
\end{table}
The value of eight parameters are summarized in Table \ref{tab:mass fit para}.
\begin{table*}[tbp]
\caption{ Values of 
parameters to include the 
effect of heavy quark spin symmetry breaking 
in Eqs.(\ref{eq:mass fit 0meson})-(\ref{eq:mass fit 32baryon}).}
\begin{center}
\begin{tabular}{cccccccc}\hline
	$a[\textrm{GeV}^2]$ & $b[\textrm{GeV}^2]$ & $c[\textrm{GeV}^2]$ & $d[\textrm{GeV}^2]$ &
	$w[\textrm{GeV}^3]$ & $x[\textrm{GeV}^3]$ & $y[\textrm{GeV}^3]$ & $z[\textrm{GeV}^3]$ \\ 
	-2.0798 & -1.8685 & 1.9889 & 2.0814 & 
	2.9468 & 3.1677 & -3.1729 & -3.0629 \\ \hline
\end{tabular}
\label{tab:mass fit para}
\end{center}
\end{table*}
When $\mu = 1.102$ and $2.779$ GeV, 
the charmed and the bottomed hadron masses are reproduced, respectively.
The heavy quark spin symmetry restores as the mass parameter $\mu$ increases.

\begin{figure}[!tbp]
\begin{center}
\includegraphics[bb = 0 0 640 384, width=0.45\textwidth]{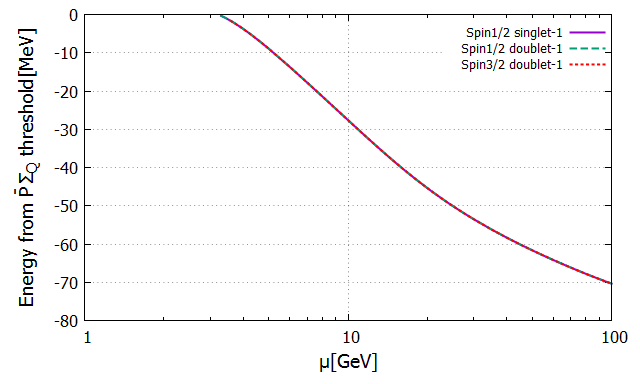}
\includegraphics[bb = 0 0 640 384, width=0.45\textwidth]{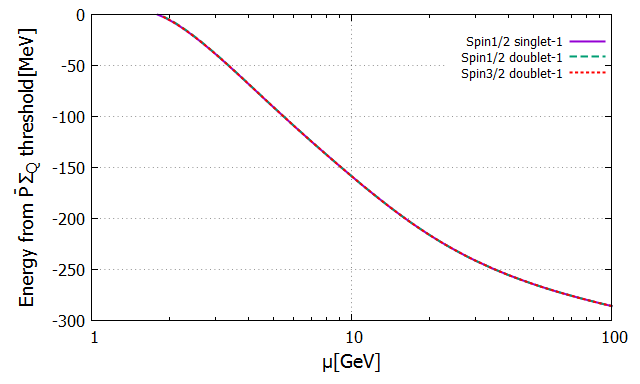}
\caption{Energies for singlet-1 and doublet-1 which have the light-cloud
 spin structure of
 $\left[ P\left[q\left[ d \right]_1\right]_{1/2} \right]_{1/2}$ with $g = +0.59$. 
The cutoff parameter is $\Lambda=1000$ MeV (upper figure) and $\Lambda=1500$ MeV (lower figure)~.
The energy is measured from the threshold of $\bar{P}\Sigma_{Q}$.
Reduced mass parameter is changed from $1$ GeV to $100$ GeV.
The labels are defined by the main component at heavy quark limit.
}
\label{fig:LCS12_12_gplus}
\end{center}
\end{figure}

\begin{figure}[!tbp]
\begin{center}
\includegraphics[bb = 0 0 640 384, width=0.45\textwidth]{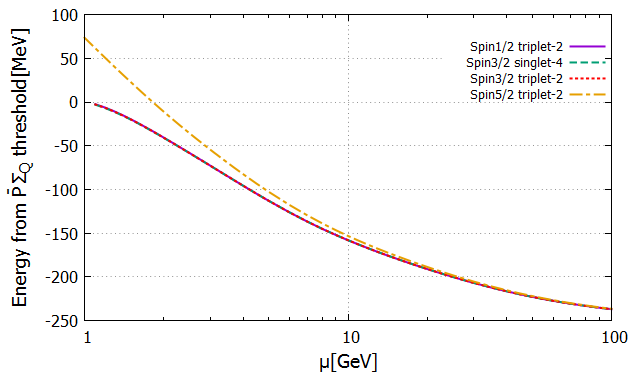}
\includegraphics[bb = 0 0 640 384, width=0.45\textwidth]{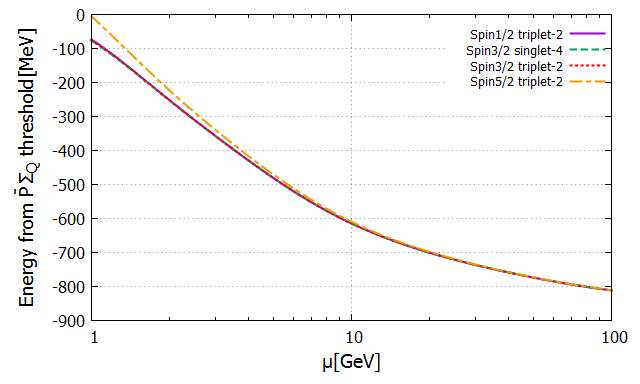}
\caption{Energies for singlet-4 and triplet-2 which have the light-cloud
 spin structure of
 $\left[ P\left[q\left[ d \right]_1\right]_{3/2} \right]_{3/2}$ with $g = +0.59$. 
The cutoff parameter is $\Lambda=1000$ MeV (upper figure) and $\Lambda=1500$ MeV (lower figure)~.
}
\label{fig:LCS32_32_gplus}
\end{center}
\end{figure}

\begin{figure}[!htbp]
\begin{center}
\includegraphics[bb = 0 0 640 384, width=0.45\textwidth]{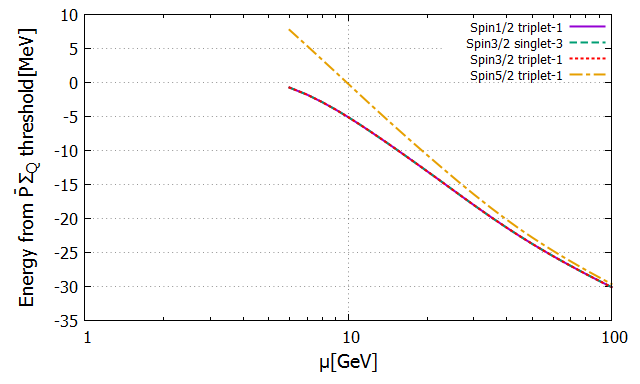}
\includegraphics[bb = 0 0 640 384, width=0.45\textwidth]{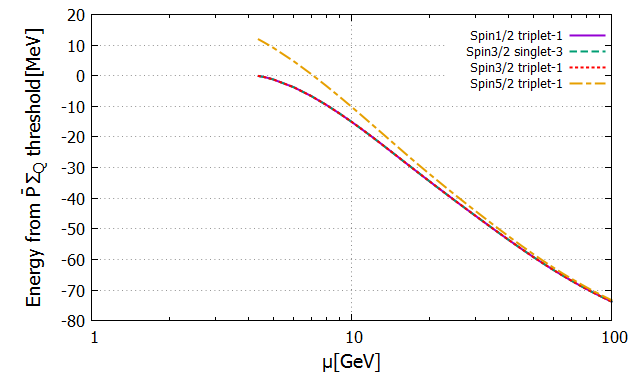}
\caption{Energies for singlet-3 and triplet-1 which have the light-cloud
 spin structure of
 $\left[ P\left[q\left[ d \right]_1\right]_{1/2} \right]_{3/2}$ with $g = -0.59$. 
The cutoff parameter is $\Lambda=1000$ MeV (upper figure) and $\Lambda=1500$ MeV (lower figure)~.
}
\label{fig:LCS12_32_gminus}
\end{center}
\end{figure}

\begin{figure}[!htbp]
\begin{center}
\includegraphics[bb = 0 0 640 384, width=0.45\textwidth]{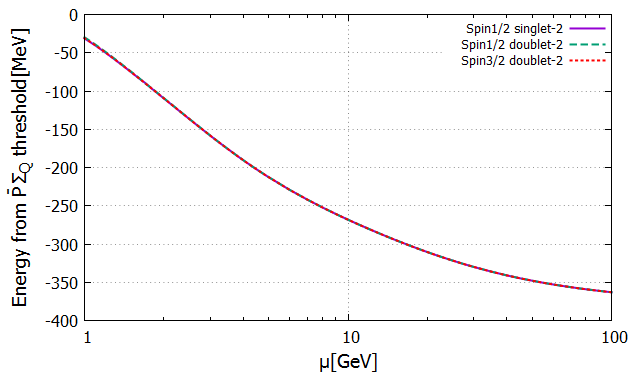}
\includegraphics[bb = 0 0 640 384, width=0.45\textwidth]{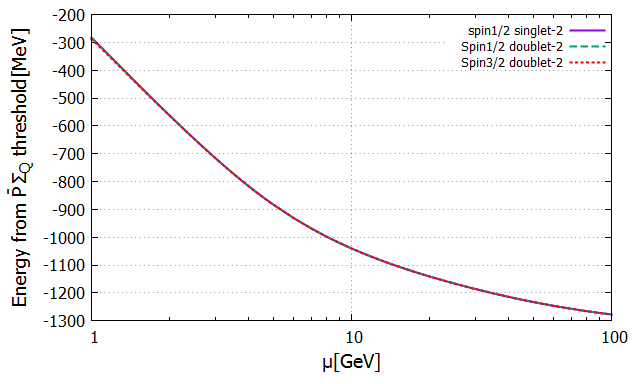}
\caption{Energies for singlet-2 and doublet-2 which have the light-cloud
 spin structure of
 $\left[ P\left[q\left[ d \right]_1\right]_{3/2} \right]_{1/2}$ with $g = -0.59$. 
The cutoff parameter is $\Lambda=1000$ MeV (upper figure) and $\Lambda=1500$ MeV (lower figure)~.
}
\label{fig:LCS32_12_gminus}
\end{center}
\end{figure}

\begin{figure}[!tbp]
\begin{center}
\includegraphics[bb = 0 0 640 384, width=0.45\textwidth]{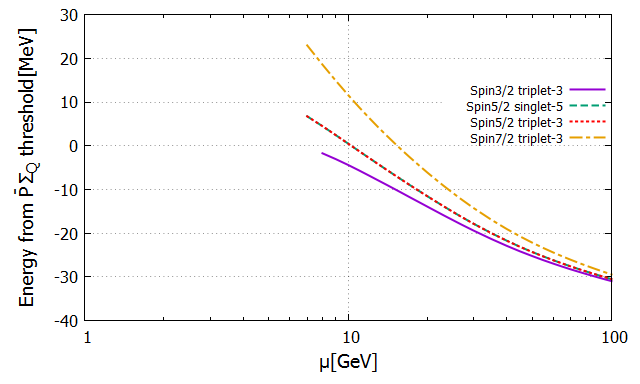}
\includegraphics[bb = 0 0 640 384, width=0.45\textwidth]{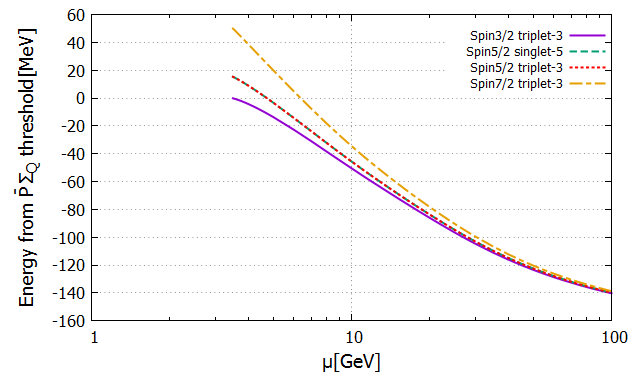}
\caption{Energies for singlet-5 and triplet-3 which have the light-cloud
 spin structure of
 $\left[ P\left[q\left[ d \right]_1\right]_{3/2} \right]_{5/2}$ with $g = -0.59$. 
The cutoff parameter is $\Lambda=1000$ MeV (upper figure) and $\Lambda=1500$ MeV (lower figure)~.
}
\label{fig:LCS32_52_gminus}
\end{center}
\end{figure}

Firstly, we show the numerical results 
obtained by solving the coupled channel Schr\"odinger equations
for each structure of light-cloud spin
in the case of $g=+0.59$.
In this case, the multiplets which have the light-clouds 
$\left[ P\left[q\left[ d \right]_1\right]_{1/2} \right]_{1/2}$ and 
$\left[ P\left[q\left[ d \right]_1\right]_{3/2} \right]_{3/2}$
are attractive.
In Fig.\ref{fig:LCS12_12_gplus}, we show the 
energy of
$\left[ P\left[q\left[ d \right]_1\right]_{1/2} \right]_{1/2}$, which corresponds to the 
spin $1/2$ singlet (s-1) in Eq.~(\ref{LCS s-1}) and spin $(1/2, 3/2)$
doublet (d-1) in Eq.~(\ref{LCS d-1}).
The labels in Fig.~\ref{fig:LCS12_12_gplus} to Fig.~\ref{fig:LCS32_52_gminus} are named by a main component of the wave function at the heavy quark limit.
Here, all the energies for three states are measured from the lowest threshold of $\bar{P}\Sigma_Q$.
Their energies are almost degenerate at whole range of $\mu$.
Next, in Fig. \ref{fig:LCS32_32_gplus}, we show the energy of
$\left[ P\left[q\left[ d \right]_1\right]_{3/2} \right]_{3/2}$, 
which corresponds to the spin $3/2$ singlet (s-4) in Eq.~(\ref{LCS s-4}) 
and spin $(1/2, 3/2, 5/2)$ triplet (t-2) in Eq.~(\ref{LCS t-2}).
We note that the lowest threthold of the spin $1/2$ and $3/2$ states is $\bar{P}\Sigma_Q$, while that of the spin $5/2$ state is $\bar{P}\Sigma_Q^*$.
Then, the energies of spin-$5/2$ state shown in Fig.~\ref{fig:LCS32_32_gplus} (and Figs.~\ref{fig:LCS12_32_gminus} and \ref{fig:LCS32_52_gminus}) are positive even if they are bound states. 
For example, in the result of $\Lambda=1000$ MeV, the mass of spin $5/2$ state is $4381.6$ MeV at $\mu=1.102$ GeV.
This value is very close to the mass of $P_c^+(4380)$.

Next, we show the result of $g=-0.59$~.
The multiplets with 
$\left[ P\left[q\left[ d \right]_1\right]_{1/2} \right]_{3/2}$, 
$\left[ P\left[q\left[ d \right]_1\right]_{3/2} \right]_{1/2}$, and
$\left[ P\left[q\left[ d \right]_1\right]_{3/2} \right]_{5/2}$ 
are attractive.
We show the energy of
$\left[ P\left[q\left[ d \right]_1\right]_{1/2} \right]_{3/2}$ in Fig. \ref{fig:LCS12_32_gminus}, 
which corresponds to the spin $3/2$ singlet (s-3) in Eq.(\ref{LCS s-3}) 
and spin $(1/2, 3/2, 5/2)$ triplet (t-1) in Eq.(\ref{LCS t-1}).
The lowest threthold of the spin $1/2$ and $3/2$ states is $\bar{P}\Sigma_Q$,
while the spin $5/2$ state is $\bar{P}\Sigma_Q^*$.
Next, 
the energy of $\left[ P\left[q\left[ d \right]_1\right]_{3/2} \right]_{1/2}$ is 
shown in Fig. \ref{fig:LCS32_12_gminus},
which corresponds to 
the spin $1/2$ singlet (s-2) in Eq.(\ref{LCS s-2}) 
and spin $(1/2, 3/2)$ doublet (d-2) in Eq.(\ref{LCS d-2}).
Finally,
we show the energy of $\left[ P\left[q\left[ d \right]_1\right]_{3/2} \right]_{5/2}$ in Fig. \ref{fig:LCS32_52_gminus}, which corresponds to the
spin $5/2$ singlet (s-5) 
in Eq.(\ref{LCS s-5})
and spin $(3/2, 5/2, 7/2)$ triplet (t-3) 
in Eq.(\ref{LCS t-3}).
We note that the threshold of the spin $7/2$ state is measured from $\bar{P}^*\Sigma_Q^*$~.
The difference of threshold values between $\bar{P}\Sigma_Q$ and $\bar{P}\Sigma_Q^*$ is $73.774$ MeV,
and between $\bar{P}\Sigma_Q$ and $\bar{P}^*\Sigma_Q^*$ is $234.63$ MeV at $\mu = 1.0$ GeV.

The states which have same structure of light-cloud spin are degenerate at heavy quark limit.
However, the heavy quark spin symmetry is broken for finite quark mass, and the mass degeneracy is resolved.
We summarize the values of the masses of obtained bound states 
at the charm region ($\mu=1.102$ GeV) and bottom region ($\mu=2.779$ GeV)
in Table \ref{tab:mass_L1000_gplus}-\ref{tab:mass_L1500_gminus}.
For instance, in Table. \ref{tab:mass_L1000_gplus}, the mass difference between $J^P=3/2^+$ and $5/2^+$ 
in HQS triplet-3 is $64.2$ MeV at $\mu=1.102$ GeV, and $24$ MeV at $\mu=2.779$ GeV.
This shows that the mass difference becomes smaller in the bottom sector.
Although the present study includes only OPEP, 
the mass degeneracy should occur even using the more realistic potential model 
if the heavy quark spin symmetry exists in the doubly heavy hadronic molecule.
The search for the HQS partner states of $P_c$-like pentaquark is very important 
to understand the structure of heavy hadrons.
\begin{table*}[!tbp]
\begin{center}
\caption{
Masses of bound states in unit of MeV with $\Lambda=1000$ MeV and $g=+0.59$ at $\mu = 1.102$ and $2.779$ GeV.}
\begin{tabular}{c|ccc|ccc|c}\hline
$\mu$ [GeV] &  & Spin $1/2$ &  &  & Spin $3/2$ &  & Spin $5/2$ \\
 & singlet-1 & doublet-1 & triplet-2 & singlet-4 & doublet-1 & triplet-2 & triplet-2 \\ \hline
$1.102$ & no bound & no bound & 4317.7 & 4317.1 & no bound & 4317.4 & 4381.6 \\
$2.779$ & no bound & no bound & 11025 & 11025 & no bound & 11025 & 11049 \\ \hline
\end{tabular}
\label{tab:mass_L1000_gplus}
\end{center}
\end{table*}
\begin{table*}[!tbp]
\begin{center}
\caption{
Masses of bound states in unit of MeV with $\Lambda=1500$ MeV and $g=+0.59$ at $\mu = 1.102$ and $2.779$ GeV.}
\begin{tabular}{c|ccc|ccc|c}\hline
$\mu$ [GeV] &  & Spin $1/2$ &  &  & Spin $3/2$ &  & Spin $5/2$ \\
 & singlet-1 & doublet-1 & triplet-2 & singlet-4 & doublet-1 & triplet-2 & triplet-2 \\ \hline
$1.102$ & no bound & no bound & 4225.4 & 4222.3 & no bound & 4223.7 & 4285.3 \\
$2.779$ & 11060 & 11060 & 10752 & 10752 & 11060 & 10752 & 11060 \\ \hline
\end{tabular}
\label{tab:mass_L1500_gplus}
\end{center}
\end{table*}
\begin{table*}[!tbp]
\begin{center}
\caption{
Masses of bound states in unit of MeV with $\Lambda=1000$ MeV and $g=-0.59$ at $\mu = 1.102$ and $2.779$ GeV.}
\begin{tabular}{c|ccc|cccc}\hline
$\mu$ [GeV] &  & Spin $1/2$ &  &  & Spin $3/2$ &  &  \\
 & singlet-2 & doublet-2 & triplet-1 & singlet-3 & doublet-2 & triplet-1 & triplet-3 \\ \hline
$1.102$ & 4279.7 & 4281.0 & no bound & no bound & 4280.0 & no bound & no bound \\
$2.779$ & 10947 & 10947 & no bound & no bound & 10942 & no bound & no bound \\ \hline \hline
 &  & Spin $5/2$ &  & & Spin7/2 \\
$\mu$ [GeV] & singlet-5 & triplet-1 & triplet-3 & & triplet-3 \\ \hline
$1.102$ & no bound & no bound & no bound & & no bound \\
$2.779$ & no bound & no bound & no bound & & no bound \\ \hline
\end{tabular}
\label{tab:mass_L1000_gminus}
\end{center}
\end{table*}
\begin{table*}[!tbp]
\begin{center}
\caption{
Masses of bound states in unit of MeV with $\Lambda=1500$ MeV and $g=-0.59$ at $\mu = 1.102$ and $2.779$ GeV.}
\begin{tabular}{c|ccc|cccc}\hline
$\mu$ [GeV] &  & Spin $1/2$ &  &  & Spin $3/2$ &  &  \\
 & singlet-2 & doublet-2 & triplet-1 & singlet-3 & doublet-2 & triplet-1 & triplet-3 \\ \hline
$1.102$ & 3993.9 & 3998.7 & no bound & no bound & 3994.9 & no bound & no bound \\
$2.779$ & 10401 & 10401 & no bound & no bound & 10401 & no bound & no bound \\ \hline \hline
 &  & Spin $5/2$ &  & & Spin7/2 \\
$\mu$ [GeV] & singlet-5 & triplet-1 & triplet-3 & & triplet-3 \\ \hline
$1.102$ & no bound & no bound & no bound & & no bound \\
$2.779$ & no bound & no bound & no bound & & no bound \\ \hline
\end{tabular}
\label{tab:mass_L1500_gminus}
\end{center}
\end{table*}


\section{Summary and Discussions}
\label{sec:summary}
In Sec.\ref{sec:HQS multiplet Pwave}, 
we showed the HQS multiplet structure of 
molecular states made from a heavy meson and a heavy baryon in P-wave.
There are five HQS singlets, two doublets, and three triplets.
The potential matrix is block diagonalized in the LCS basis for each HQS multiplet as shown in Sec.~\ref{sec:Lagrangian and Potential}.
We obtained the binding energy by solving the Schr\"odinger equation under OPEP in Sec.~\ref{sec:Numerical result}.
When $g$ is positive, the spin $1/2$ singlet (s-1), $3/2$ singlet (s-4), 
the blocks of the potential for $(1/2, 3/2)$ doublet (d-1), and $(1/2, 3/2, 5/2)$ triplet (t-2) are attractive.
The blocks for other six multiplets, $1/2$ singlet (s-2), $3/2$ singlet (s-3), $5/2$ singlet (s-5), 
$(1/2, 3/2)$ doublet (d-2), $(1/2, 3/2, 5/2)$ triplet (t-1), and $(3/2, 5/2, 7/2)$ triplet (t-3) are attractive when $g$ is negative.
The behavior of the binding energy is classified by the structure of the light-cloud spin.
As mentioned in Ref.\cite{Shimizu:2018ran}, 
OPEP depends only on the structure of the light-cloud spin
since the pion exchange interaction couples the light quark spin and the orbital angular momentum.
The HQS multiplets having the same light-cloud structure are degenerate at heavy quark limit.
The mass degeneracy is resolved for hidden-charm/bottom pentaquarks because of the finite quark mass.

The mass of the $J^P=5/2^+$ state in Tab.~\ref{tab:mass_L1000_gplus} is close to the one of $P_c^+(4380)$.
It's HQS triplet partner states carrying $J^P=1/2^+$ and $3/2^+$ exist around $4320$ MeV.
The masses of their hidden-bottom flavor partner are $11025$ MeV for $J^P=1/2^+$ and $3/2^+$ states, and $11049$ MeV for $5/2^+$ states.
We expect that these partner states will be found in future experiments.

When $P_c^+(4380)$ is $J^P=5/2^+$, $P_c^+(4450)$ should be $3/2^-$.
We did not obtain such state in our previous work \cite{Shimizu:2018ran},
since the mass of $P_c^+(4450)$ is above the threshold of our coupled channel.
The full coupled channel analysis of $\bar{D}^{(*)}\Lambda_c-\bar{D}^{(*)}\Sigma_c^{(*)}$
using the complex scaling method was done in Ref.\cite{Yamaguchi:2016ote}, 
which shows that it seems to be difficult to explain the mass and decay width of both $P_c(4380)$ and $P_c(4450)$ at the same time.
Moreover, $P_c$ pentaquarks are not reproduced 
by the estimation of compact five-body pentaquark \cite{Hiyama:2018ukv}.
Some works argue that it is a threshold cusp by kinematical effect \cite{Guo:2015umn, Liu:2015fea, Guo:2016bkl, Bayar:2016ftu}.
More theoretical efforts and experimental data are needed to reveal the nature of the $P_c$ pentaquarks.

\begin{figure}[!htbp]
\begin{center}
\includegraphics[bb = 0 0 1000 560, width=0.47\textwidth]{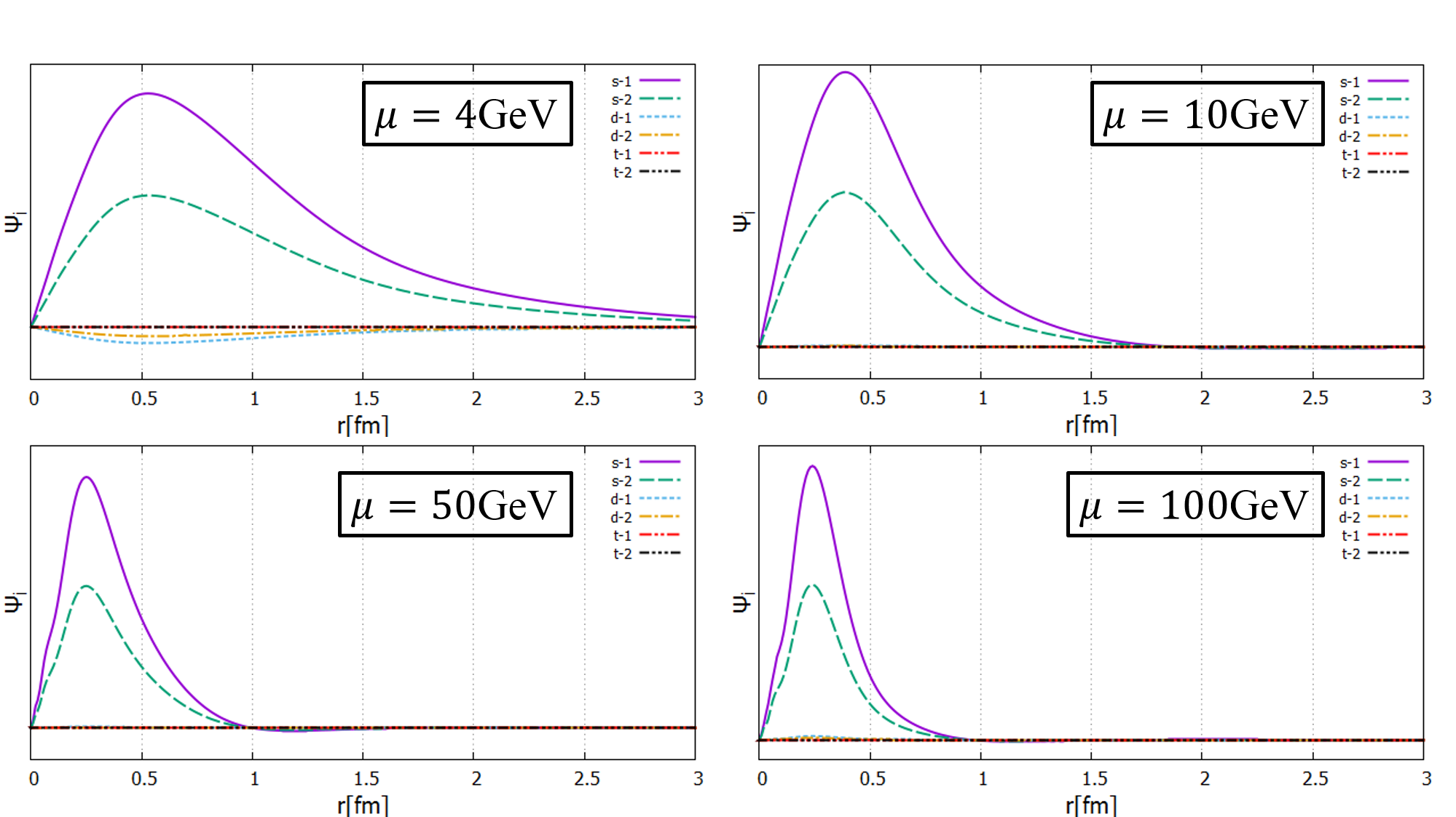}
\caption{
Wave functions of singlet-1 state with $J^P=1/2^+$, $g=+0.59$ and $\Lambda=1000$ MeV.
The case of four different mass parameters
are shown.
}
\label{fig:wf_12_gplus_s1}
\end{center}
\end{figure}

\begin{figure}[!htbp]
\begin{center}
\includegraphics[bb = 0 0 960 540, width=0.47\textwidth]{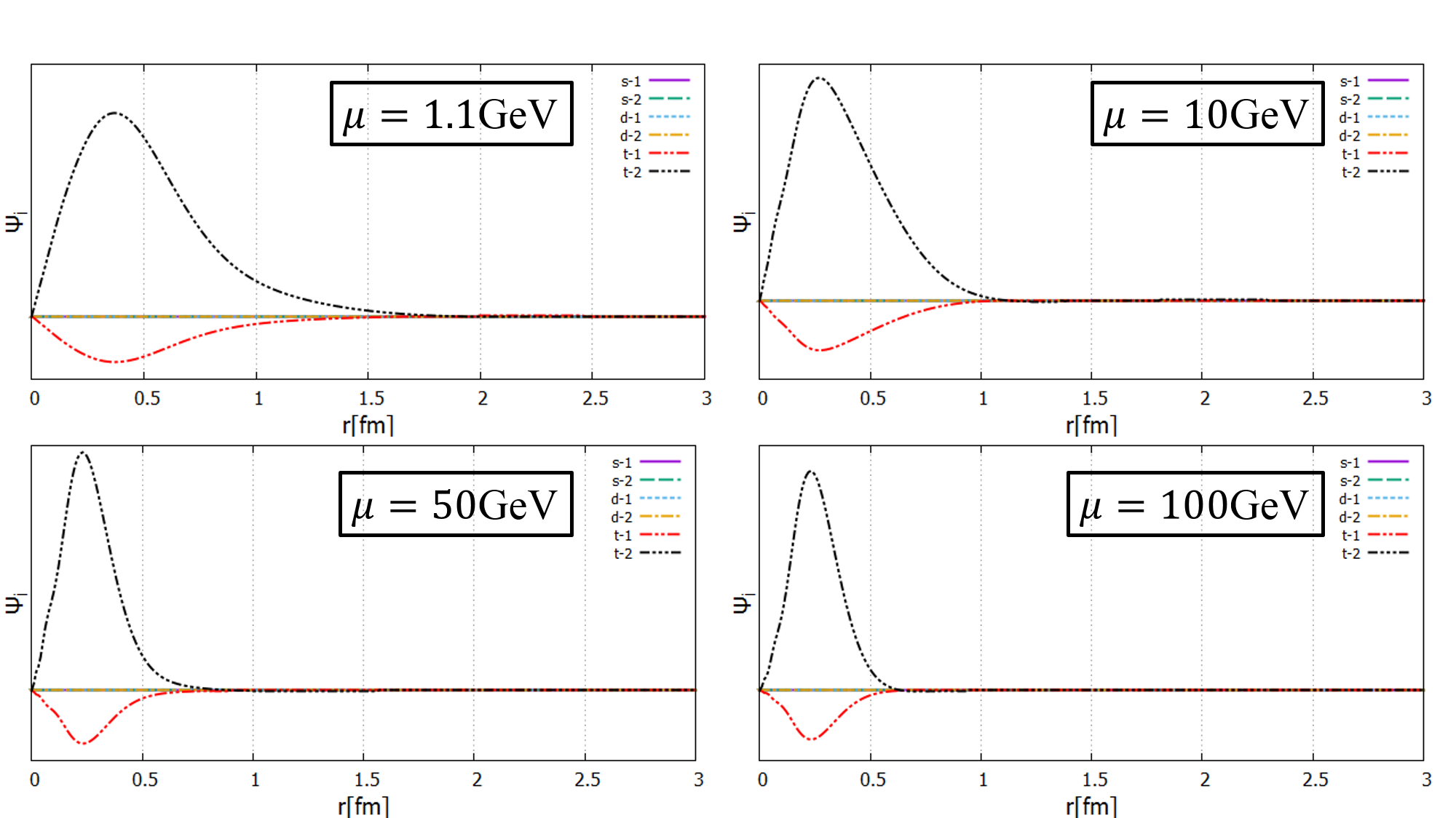}
\caption{
Wave functions of triplet-2 state with $J^P=1/2^+$, $g=+0.59$ and $\Lambda=1000$ MeV.
The case of four different mass parameters
are shown.
}
\label{fig:wf_12_gplus_t2}
\end{center}
\end{figure}

We determined the names of the solutions by the dominant component of
the wave function,
which we obtain together with the binding energy when solving the Schr\"odinger equation.
For instance, in Fig.\ref{fig:wf_12_gplus_s1} and Fig.\ref{fig:wf_12_gplus_t2}, 
we show the obtained wave functions of singlet-1 and triplet-2 with $J^P=1/2^+$,
for which the binding energies are shown in Fig.\ref{fig:LCS12_12_gplus} and Fig.\ref{fig:LCS32_32_gplus}, respectively.
We note that these wave functions are not normalized, therefore only the ratio of components is meaningful.
In both cases, the change in the ratio of the wave functions when changing the mass parameter is very small.
In the case of the finite quark mass, although the different components of the HQS multiplet are mixed, the ratio is still small.
This shows that the effect of the symmetry breaking by the kinetic terms is small.
When the mass parameter becomes larger, the wave functions concentrate at a position where the tensor potential becomes deep.
This means that as the mass increases, the kinetic term is suppressed and the wave function is localized at the bottom of the attractive potential.

To treat the heavy quark spin symmetry for doubly heavy system, 
we assume that two heavy quarks in a pentaquark are labeled by the same velocity.
This assumption seems to be natural to the bound state 
because a hadronic molecule breaks when the relative velocity between the meson and baryon is large.
However, if the heavy meson and heavy baryon are labeled by the same velocity, 
they can not get the orbital angular momentum exitation.
To include the P-wave exitation we label one hadron by $v$ and another by $v'$, and assume that their difference is small.
Namely, when $v'$ is written as $v' = v + \delta v$ with the small quantity $\delta v$,
we classify the HQS multiplet structure of P-wave states at the leading order of $v'$ by neglecting $\delta v$.
We do not include the effect of $\delta v$ in the present study and  leave the inclusion of its effects for future publications.

\acknowledgments
The work of Y.S. is supported in part by JSPS Grant-in-Aid for JSPS Research Fellow No. JP17J06300. 
The work of Y.Y. is supported in part by the Special Postdoctoral Researcher (SPDR) and iTHEMS Programs of RIKEN.
The work of M.H. is supported in part by 
JPSP KAKENHI
Grant Number 16K05345.

\appendix
\section{Basis transformation}
\label{sec:Basis transformation}
We show the detail of the transformation from HM basis to LCS basis.
The wave function transformation is done by Eq.(\ref{eq:wave function transformation}).
The component of wave functions in HM basis for each spin state is
\begin{align}
\psi_{1/2^+}^{\rm HM} &= \left(
\begin{array}{l}
\bar{P}\Sigma_Q\left(^2 P_{1/2} \right) \\
\bar{P}\Sigma_Q^*\left(^4 P_{1/2} \right) \\
\bar{P}^*\Sigma_Q\left(^2 P_{1/2} \right) \\
\bar{P}^*\Sigma_Q\left(^4 P_{1/2} \right) \\
\bar{P}^*\Sigma_Q^*\left(^2 P_{1/2} \right) \\
\bar{P}^*\Sigma_Q^*\left(^4 P_{1/2} \right)
\end{array}
\right), 
\end{align}
\begin{align}
\psi_{3/2^+}^{\rm HM} = \left(
\begin{array}{c}
\bar{P}\Sigma_Q\left(^2 P_{3/2} \right) \\
\bar{P}\Sigma_Q^*\left(^4 P_{3/2} \right) \\
\bar{P}^*\Sigma_Q\left(^2 P_{3/2} \right) \\
\bar{P}^*\Sigma_Q\left(^4 P_{3/2} \right) \\
\bar{P}^*\Sigma_Q^*\left(^2 P_{3/2} \right) \\
\bar{P}^*\Sigma_Q^*\left(^4 P_{3/2} \right) \\
\bar{P}^*\Sigma_Q^*\left(^6 P_{3/2} \right)
\end{array}
\right), 
\end{align}
\begin{align}
\psi_{5/2^+}^{\rm HM} = \left(
\begin{array}{c}
\bar{P}\Sigma_Q^*\left(^4 P_{5/2} \right) \\
\bar{P}^*\Sigma_Q\left(^4 P_{5/2} \right) \\
\bar{P}^*\Sigma_Q^*\left(^4 P_{5/2} \right) \\
\bar{P}^*\Sigma_Q^*\left(^6 P_{5/2} \right)
\end{array}
\right), 
\end{align}
\begin{align}
\psi_{7/2^+}^{\rm HM} = \left(
\begin{array}{c}
\bar{P}^*\Sigma_Q^*\left(^6 P_{7/2} \right)
\end{array}
\right). 
\end{align}
The basis transformation is done by
\begin{align}
\psi_{J^P}^{\rm LCS} = U_{J^P}^{-1}\psi_{J^P}^{\rm HM}.
\end{align}
The component of wave functions in LCS basis is
\begin{align}
\psi_{1/2^+}^{\rm LCS} = \left(
\begin{array}{c}
\left[ \left[\bar{Q}Q\right]_0 \left[ P\left[q\left[d\right]_1\right]_{1/2} \right]_{1/2} \right]_{1/2}^{\rm singlet-1} \\
\left[ \left[\bar{Q}Q\right]_0 \left[ P\left[q\left[d\right]_1\right]_{3/2} \right]_{1/2} \right]_{1/2}^{\rm singlet-2} \\
\left[ \left[\bar{Q}Q\right]_1 \left[ P\left[q\left[d\right]_1\right]_{1/2} \right]_{1/2} \right]_{1/2}^{\rm doublet-1} \\
\left[ \left[\bar{Q}Q\right]_1 \left[ P\left[q\left[d\right]_1\right]_{3/2} \right]_{1/2} \right]_{1/2}^{\rm doublet-2} \\
\left[ \left[\bar{Q}Q\right]_1 \left[ P\left[q\left[d\right]_1\right]_{1/2} \right]_{3/2} \right]_{1/2}^{\rm triplet-1} \\
\left[ \left[\bar{Q}Q\right]_1 \left[ P\left[q\left[d\right]_1\right]_{3/2} \right]_{3/2} \right]_{1/2}^{\rm triplet-2} 
\end{array}
\right),
\label{eq:wave function in LCS 1/2}
\end{align}
\begin{align}
\psi_{3/2^+}^{\rm LCS} = \left(
\begin{array}{c}
\left[ \left[\bar{Q}Q\right]_0 \left[ P\left[q\left[d\right]_1\right]_{1/2} \right]_{3/2} \right]_{3/2}^{\rm singlet-3} \\
\left[ \left[\bar{Q}Q\right]_0 \left[ P\left[q\left[d\right]_1\right]_{3/2} \right]_{3/2} \right]_{3/2}^{\rm singlet-4} \\
\left[ \left[\bar{Q}Q\right]_1 \left[ P\left[q\left[d\right]_1\right]_{1/2} \right]_{1/2} \right]_{3/2}^{\rm doublet-1} \\
\left[ \left[\bar{Q}Q\right]_1 \left[ P\left[q\left[d\right]_1\right]_{3/2} \right]_{1/2} \right]_{3/2}^{\rm doublet-2} \\
\left[ \left[\bar{Q}Q\right]_1 \left[ P\left[q\left[d\right]_1\right]_{1/2} \right]_{3/2} \right]_{3/2}^{\rm triplet-1} \\
\left[ \left[\bar{Q}Q\right]_1 \left[ P\left[q\left[d\right]_1\right]_{3/2} \right]_{3/2} \right]_{3/2}^{\rm triplet-2} \\
\left[ \left[\bar{Q}Q\right]_1 \left[ P\left[q\left[d\right]_1\right]_{3/2} \right]_{5/2} \right]_{3/2}^{\rm triplet-3} 
\end{array}
\right), 
\label{eq:wave function in LCS 3/2}
\end{align}
\begin{align}
\psi_{5/2^+}^{\rm LCS} = \left(
\begin{array}{c}
\left[ \left[\bar{Q}Q\right]_0 \left[ P\left[q\left[d\right]_1\right]_{3/2} \right]_{5/2} \right]_{5/2}^{\rm singlet-5}  \\
\left[ \left[\bar{Q}Q\right]_1 \left[ P\left[q\left[d\right]_1\right]_{1/2} \right]_{3/2} \right]_{5/2}^{\rm triplet-1}  \\
\left[ \left[\bar{Q}Q\right]_1 \left[ P\left[q\left[d\right]_1\right]_{3/2} \right]_{3/2} \right]_{5/2}^{\rm triplet-2}  \\
\left[ \left[\bar{Q}Q\right]_1 \left[ P\left[q\left[d\right]_1\right]_{3/2} \right]_{5/2} \right]_{5/2}^{\rm triplet-3} 
\end{array}
\right), 
\label{eq:wave function in LCS 5/2}
\end{align}
\begin{align}
\psi^{\rm LCS}_{7/2^+} &= \left(
\begin{array}{c}
\left[ \left[\bar{Q}Q\right]_1 \left[ P\left[q\left[d\right]_1\right]_{3/2} \right]_{5/2} \right]_{7/2}^{\rm triplet-3} 
\end{array}
\right).
\label{eq:wave function in LCS 7/2}
\end{align}
The notation of spin structure is same with in Sec.\ref{sec:HQS multiplet Pwave}.
The transformation matrix $U$ is determined by the Clebsch-Gordan coefficient 
to reconstruct the spin structures from HM basis to LCS basis.
\begin{widetext}
\begin{align}
U_{1/2^+} &= \left(
\begin{array}{cccccc}\vspace{2pt}
\frac{1}{2} & 0 & \frac{\sqrt{3}}{18} & \frac{2\sqrt{6}}{9} & \frac{\sqrt{6}}{9} & \frac{\sqrt{30}}{9} \\ \vspace{2pt}
0 & \frac{1}{2} & \frac{2\sqrt{6}}{9} & \frac{5\sqrt{3}}{18} & -\frac{\sqrt{3}}{9} & -\frac{\sqrt{15}}{9} \\ \vspace{2pt}
-\frac{\sqrt{3}}{6} & 0 & -\frac{5}{18} & \frac{2\sqrt{2}}{9} & -\frac{5\sqrt{2}}{9} & \frac{\sqrt{10}}{9} \\ \vspace{2pt}
0 & -\frac{\sqrt{3}}{3} & -\frac{2\sqrt{2}}{9} & \frac{5}{9} & \frac{1}{9} & -\frac{2\sqrt{5}}{9} \\ \vspace{2pt}
\frac{\sqrt{6}}{3} & 0 & -\frac{\sqrt{2}}{9} & -\frac{2}{9} & -\frac{4}{9} & -\frac{\sqrt{5}}{9} \\ \vspace{2pt}
0 & \frac{\sqrt{15}}{6} & -\frac{2\sqrt{10}}{9} & \frac{\sqrt{5}}{18} & \frac{\sqrt{5}}{9} & -\frac{1}{9} 	
\end{array}
\right), 
\end{align}
\begin{align}
U_{3/2^+} = \left(
\begin{array}{ccccccc}\vspace{2pt}
\frac{1}{2} & 0 & -\frac{\sqrt{3}}{9} & \frac{\sqrt{6}}{18} & -\frac{\sqrt{15}}{18} & \frac{2\sqrt{3}}{9} & \frac{\sqrt{2}}{2} \\ \vspace{2pt}
0 & \frac{1}{2} & -\frac{\sqrt{15}}{9} & \frac{\sqrt{30}}{18} & \frac{2\sqrt{3}}{9} & \frac{11\sqrt{15}}{90} & -\frac{\sqrt{10}}{10} \\ \vspace{2pt}
-\frac{\sqrt{3}}{6} & 0 & \frac{5}{9} & \frac{\sqrt{2}}{18} & \frac{5\sqrt{5}}{18} & \frac{2}{9} & \frac{\sqrt{6}}{6} \\ \vspace{2pt}
0 & -\frac{\sqrt{3}}{3} & \frac{\sqrt{5}}{9} & \frac{\sqrt{10}}{9} & -\frac{2}{9} & \frac{11\sqrt{5}}{45} & -\frac{\sqrt{30}}{15} \\ \vspace{2pt}
\frac{\sqrt{6}}{3} & 0 & \frac{2\sqrt{2}}{9} & -\frac{1}{18} & \frac{\sqrt{10}}{9} & -\frac{\sqrt{2}}{9} & -\frac{\sqrt{3}}{6} \\ \vspace{2pt}
0 & \frac{\sqrt{15}}{6} & \frac{5}{9} & \frac{\sqrt{2}}{18} & -\frac{2\sqrt{5}}{9} & \frac{11}{90} & -\frac{\sqrt{6}}{30} \\ \vspace{2pt}
0 & 0 & 0 & \frac{\sqrt{3}}{2} & 0 & -\frac{\sqrt{6}}{5} & \frac{1}{10}
\end{array}
\right), 
\end{align}
\end{widetext}
\begin{align} 
U_{5/2^+} &= \left(
\begin{array}{cccc}
\frac{1}{2} & -\frac{\sqrt{3}}{3} & \frac{\sqrt{15}}{15} & \frac{\sqrt{35}}{10} \\
-\frac{\sqrt{3}}{3} & \frac{1}{3} & \frac{2\sqrt{5}}{15} & \frac{\sqrt{105}}{15} \\
\frac{\sqrt{15}}{6} & \frac{\sqrt{5}}{3} & \frac{1}{15} & \frac{\sqrt{21}}{30} \\
0 & 0 & \frac{\sqrt{21}}{5} & -\frac{2}{5}
\end{array}
\right), 
\end{align}
\begin{align}
U_{7/2^+} = 1~.
\end{align}

In Sec.\ref{sec:Lagrangian and Potential} we construct the one-pion exchange potential 
from the heavy hadron effective Lagrangians in HM basis.
The detail of the OPEP matrix in HM basis is as follows : 
\begin{widetext}
\begin{align}
V_{1/2^+}^{\rm HM} &= \left(
\begin{array}{cccccc}
	0 & 0 & -\frac{\sqrt{3}}{3}C & \frac{\sqrt{6}}{3}T & \frac{\sqrt{6}}{6}C & \frac{\sqrt{30}}{30}T \vspace{2pt} \\
	0 & 0 & \frac{\sqrt{6}}{6}T & -\frac{\sqrt{3}}{6}C -\frac{\sqrt{3}}{6}T & -\frac{\sqrt{3}}{6}T & -\frac{\sqrt{15}}{6}C +\frac{2\sqrt{15}}{15}T \vspace{2pt} \\
	-\frac{\sqrt{3}}{3}C & \frac{\sqrt{6}}{6}T & \frac{2}{3}C & \frac{\sqrt{2}}{3}T & \frac{\sqrt{2}}{6}C & \frac{2\sqrt{10}}{15}T \vspace{2pt} \\
	\frac{\sqrt{6}}{3}T & -\frac{\sqrt{3}}{6}C - \frac{\sqrt{3}}{6}T & \frac{\sqrt{2}}{3}T & -\frac{1}{3}C +\frac{2}{3}T & \frac{1}{6}T & \frac{\sqrt{5}}{6}C -\frac{\sqrt{5}}{30}T \vspace{2pt} \\
	\frac{\sqrt{6}}{6}C & -\frac{\sqrt{3}}{6}T & \frac{\sqrt{2}}{6}C & \frac{1}{6}T & \frac{5}{6}C & -\frac{7\sqrt{5}}{30}T \vspace{2pt} \\
	\frac{\sqrt{30}}{30}T & -\frac{\sqrt{15}}{6}C + \frac{2\sqrt{15}}{15}T & \frac{2\sqrt{10}}{15}T & \frac{\sqrt{5}}{6}C - \frac{\sqrt{5}}{30}T & -\frac{7\sqrt{5}}{30}T & \frac{1}{3}C + \frac{8}{15}T
\end{array}
\right)\frac{gg_1}{f_{\pi}^2}~,
\end{align}
{\small
\begin{align}
V_{3/2^+}^{\rm HM} = \left(
	\begin{array}{ccccccc}
		0 & 0 & -\frac{\sqrt{3}}{3}C & -\frac{\sqrt{15}}{15}T & \frac{\sqrt{6}}{6}C & -\frac{\sqrt{3}}{30}T & \frac{3\sqrt{2}}{10}T \\
		0 & 0 & -\frac{\sqrt{15}}{30}T & -\frac{\sqrt{3}}{6}C+\frac{2\sqrt{3}}{15}T & \frac{\sqrt{30}}{60}T & -\frac{\sqrt{15}}{6}C-\frac{8\sqrt{15}}{75}T & \frac{21\sqrt{10}}{100}T \\
		-\frac{\sqrt{3}}{3}C & -\frac{\sqrt{15}}{30}T & \frac{2}{3}C & -\frac{\sqrt{5}}{15}T & \frac{\sqrt{2}}{6}C & -\frac{2}{15}T & -\frac{\sqrt{6}}{10}T \\
		-\frac{\sqrt{15}}{15}T & -\frac{\sqrt{3}}{6}C + \frac{2\sqrt{3}}{15}T & -\frac{\sqrt{5}}{15}T & -\frac{1}{3}C - \frac{8}{15}T & -\frac{\sqrt{10}}{60}T & \frac{\sqrt{5}}{6}C + \frac{2\sqrt{5}}{75}T & \frac{7\sqrt{30}}{100}T \\
		\frac{\sqrt{6}}{6}C & \frac{\sqrt{30}}{60}T & \frac{\sqrt{2}}{6}C & -\frac{\sqrt{10}}{60}T & \frac{5}{6}C & \frac{7\sqrt{2}}{60}T & -\frac{\sqrt{3}}{5}T\\
		-\frac{\sqrt{3}}{30}T & -\frac{\sqrt{15}}{6}C - \frac{8\sqrt{15}}{75}T & -\frac{2}{15}T & \frac{\sqrt{5}}{6}C + \frac{2\sqrt{5}}{75}T & \frac{7\sqrt{2}}{60}T & \frac{1}{3}C - \frac{32}{75}T & -\frac{7\sqrt{6}}{100}T \\
		\frac{3\sqrt{2}}{10}T & \frac{21\sqrt{10}}{100}T & -\frac{\sqrt{6}}{10}T & \frac{7\sqrt{30}}{100}T & -\frac{\sqrt{3}}{5}T & -\frac{7\sqrt{6}}{100}T & -\frac{1}{2}C + \frac{14}{25}T
	\end{array}
\right)\frac{gg_1}{f_{\pi}^2}~,
\end{align}
}
\begin{align}
V^{\rm HM}_{5/2^+} &= \left(
	\begin{array}{cccc}
		0 & -\frac{\sqrt{3}}{6}C - \frac{\sqrt{3}}{30}T & -\frac{\sqrt{15}}{6}C + \frac{2\sqrt{15}}{75}T & -\frac{3\sqrt{35}}{50}T \\
		-\frac{\sqrt{3}}{6}C - \frac{\sqrt{3}}{30}T & -\frac{1}{3}C + \frac{2}{15}T & \frac{\sqrt{5}}{6}C - \frac{\sqrt{5}}{150}T & -\frac{\sqrt{105}}{50}T \\
		-\frac{\sqrt{15}}{6}C + \frac{2\sqrt{15}}{75}T & \frac{\sqrt{5}}{6}C - \frac{\sqrt{5}}{150}T & \frac{1}{3}C+\frac{8}{75}T & \frac{\sqrt{21}}{50}T \\
		-\frac{3\sqrt{35}}{50}T & -\frac{\sqrt{105}}{50}T & \frac{\sqrt{21}}{50}T & -\frac{1}{2}C - \frac{16}{25}T
	\end{array}
\right)\frac{gg_1}{f_{\pi}^2}~,
\end{align}
\end{widetext}
\begin{align}
V^{\rm HM}_{7/2^+} &= \frac{gg_1}{f_{\pi}^2}\left[ -\frac{1}{2}C + \frac{1}{5}T \right]~.
\end{align}
The definition of the spin-spin potential $C$ and tensor potential $T$ 
are written in Sec.\ref{sec:Lagrangian and Potential}.
The tranformation of potential matrix from HM basis to LCS basis is done by
\begin{align}
V^{\rm LCS}_{J^P} &= U_{J^P}^{-1} V^{\rm HM}_{J^P} U_{J^P}~.
\end{align}
The potential matrices in LCS basis are written in Sec.\ref{sec:Lagrangian and Potential}.

The kinetic term is defined by
\begin{align}
K_{i}^{L} = -\frac{1}{2\mu_i}\left( \frac{\del^2}{\del r^2} + \frac{2}{r}\frac{\del}{\del r} - \frac{L(L+1)}{r^2} \right),
\end{align}
where $i$ is a channel index, $L$ is an orbital angular momentum, and $\mu_i$ is a reduced mass of channel $i$.
The kinetic term matrices for each $J^P$ in HM basis are as follows : 
\begin{align}
K_{1/2^+}^{\rm HM} &= diag\left[ K_{\bar{P}\Sigma_Q}^1, K_{\bar{P}\Sigma_Q^*}^1, K_{\bar{P}^*\Sigma_Q}^1, \right. \non \\
&\hspace{25pt}\left. K_{\bar{P}^*\Sigma_Q}^1, K_{\bar{P}^*\Sigma_Q^*}^1, K_{\bar{P}^*\Sigma_Q^*}^1 \right]~, \\
K_{3/2^+}^{\rm HM} &= diag\left[ K_{\bar{P}\Sigma_Q}^1, K_{\bar{P}\Sigma_Q^*}^1, K_{\bar{P}^*\Sigma_Q}^1, \right. \non \\
&\hspace{25pt}\left. K_{\bar{P}^*\Sigma_Q}^1, K_{\bar{P}^*\Sigma_Q^*}^1, K_{\bar{P}^*\Sigma_Q^*}^1, K_{\bar{P}^*\Sigma_Q^*}^1 \right]~, \\
K_{5/2^+}^{\rm HM} &= diag\left[ K_{\bar{P}\Sigma_Q^*}^1, K_{\bar{P}^*\Sigma_Q}^1, K_{\bar{P}^*\Sigma_Q^*}^1, K_{\bar{P}^*\Sigma_Q^*}^1 \right]~, \\
K_{7/2^+}^{\rm HM} &= K_{\bar{P}^*\Sigma_Q^*}^1~.
\end{align}
The transformation to LCS basis is done by
\begin{align}
K_{J^P}^{\rm LCS} = U_{J^P}^{-1} K_{J^P}^{\rm HM} U_{J^P}~.
\end{align}

\end{document}